\documentclass[aps]{revtex4}

\usepackage{graphicx}
\usepackage{amssymb}
\usepackage{bm}
\usepackage[linktocpage]{hyperref}
\usepackage{color}

\begin{document}

\title{Large Scale Structure Formation in Eddington-inspired Born-Infeld Gravity}

\author{Xiao-Long Du$^{1}$}
\email{duxl11@lzu.edu.cn}
\author{Ke Yang$^{1}$}
\email{yangke09@lzu.edu.cn}
\author{Xin-He Meng$^{2,3}$}
\email{xhm@nankai.edu.cn}
\author{Yu-Xiao Liu$^{1}$\footnote{Corresponding author.}}
\email{liuyx@lzu.edu.cn}

\affiliation{$^{1}$Institute of Theoretical Physics, Lanzhou University, Lanzhou 730000, P.R. China}
\affiliation{$^{2}$Department of Physics, Nankai University, Tianjin 300071, P.R. China}
\affiliation{$^{3}$State Key Laboratory of Theoretical Physics, Institute of Theoretical Physics, CAS, Beijing 100190, P.R. China\vspace{1cm}}

\begin{abstract}
We study the large scale structure formation in Eddington-inspired Born-Infeld (EiBI) gravity. It is found
that the linear growth of scalar perturbations in EiBI gravity deviates from that in general relativity
for modes with large wave numbers ($k$), but the deviation is largely suppressed with the expansion of the
Universe. We investigate the integrated Sachs-Wolfe effect in EiBI gravity, and find that its effect on the
angular power spectrum of the anisotropy of the cosmic microwave background (CMB) is almost the same as that
in the Lambda-cold dark matter ($\Lambda$CDM) model. We further calculate the linear matter power spectrum
in EiBI gravity and compare it with that in the $\Lambda$CDM model. Deviation is found on small scales
($k\gtrsim 0.1 h$ Mpc$^{-1}$), which can be tested in the future by observations from galaxy surveys.
\end{abstract}

\maketitle

\section{Introduction}\label{intro}

Purely affine theory of gravity has drawn a lot of attention since it was first proposed by Eddington \cite{Eddington1924}.
Schr\"{o}dinger generalized Eddington's theory to a nonsymmetric metric \cite{Schrodinger1950}. One of the advantages
of Eddington affine theory is that it can automatically generate a cosmological term. But in these early papers, matter fields
are not included. Attempts to add matter fields in this theory have been an interesting topic \cite{Deser:1998rj,Vollick:2003qp}.
Recently, a new alternative theory called Eddington-inspired Born-Infeld (EiBI) gravity was proposed by
Banados and Ferreira \cite{Banados:2010ix}. EiBI gravity is equivalent to general relativity in vacuum; but when matter
fields are included, it presents many interesting properties. It is claimed to be singularity free both at the
beginning of the Universe \cite{Banados:2010ix,Scargill:2012kg} and {during the gravitational collapse of dust} \cite{Pani:2011mg}.
In Ref. \cite{Avelino:2012ue}, EiBI gravity as an alternative to inflation was discussed.

Despite the good properties EiBI gravity exhibits, the validity of this theory has also been an important topic.
It was found that the tensor perturbation and the nonzero wave number modes of scalar perturbations in EiBI gravity
are unstable deep in the Eddington regime \cite{Avelino:2012ue,EscamillaRivera:2012vz,Yang:2013hsa,Lagos:2013aua},
while it was shown that the vector perturbations and the zero wave number modes of scalar perturbations are stable
for positive $\kappa$ (an extra parameter in EiBI gravity) in Ref. \cite{Yang:2013hsa}. In Ref. \cite{Pani:2012qd},
the authors argued that there exist curvature singularities at the surface of polytropic stars and unacceptable
Newtonian limit in EiBI gravity. {On the other hand, researchers try to find out how we can remove these
pathologies. In Ref. \cite{Liu:2012rc}, Liu et al. investigated a thick brane model in EiBI gravity. They found that the
instability of tensor perturbation does not exit in their model. In Ref. \cite{Avelino:2012ue}, Avelino and Ferreira found another
solution to the instability problem of tensor perturbation by considering matter sources with a time-dependent state
parameter. Recently, Kim argued that the problem of singularity at the surface of a star
can be cured by taking into account the gravitational backreaction \cite{Kim:2013nna}. These extensions make EiBI
gravity a more consistent theory and a prospective alternative to general relativity.}

Other papers have also been done to constrain the parameter $\kappa$ from compact
stars \cite{Pani:2011mg,Pani:2011xm,Sham:2013sya}, tests in solar system \cite{Casanellas:2011kf}, astrophysical
and cosmological observations \cite{Avelino:2012ge}, and nuclear physics \cite{Avelino:2012qe,Harko:2013wka}.
The strongest constraint on the parameter $\kappa$ implies $|\kappa|<10^{-3}$kg$^{-1}$m$^5$s$^{-2}$\cite{Avelino:2012qe}. More relevant studies can be found in Refs.
\cite{Pani:2012qb,Delsate:2012ky,Sham:2012qi,Cho:2012vg,Jana:2013fga,Cho:2013usa,Bouhmadi-Lopez:2013lha,Cho:2013pea,Harko:2013xma,Fiorini:2013kba,Harko:2013aya,Olmo:2013gqa,Kim:2013noa,
Sotani:2014goa,Cho:2014ija,Wei:2014dka,Sotani:2014xoa,Bouhmadi-Lopez:2014jfa,Fu:2014raa}.

It was shown in Refs. \cite{Banados:2010ix,Scargill:2012kg,Cho:2012vg} that, in the low density and curvature limit,
EiBI gravity recoveries the conventional Friedman cosmology. But these studies only considered a homogeneous and
isotropic Universe. It is worthwhile to examine the cosmological consequences of perturbations in EiBI gravity.
In Ref. \cite{Lagos:2013aua}, the authors found that a nearly scale-invariant power spectrum for both scalar and
tensor primordial quantum perturbations can be obtained in EiBI gravity without introducing the inflation.
However, it remains to be seen whether these primordial quantum perturbations can lead to proper cosmic microwave background (CMB) and
large scale structure of the Universe consistent with observations.

In this paper, we investigate the evolution of cosmological perturbations after the last scattering and the large scale
structure formation in EiBI gravity. First, we use the linear perturbed equations derived in \cite{Yang:2013hsa} to obtain
the approximate equations governing the scalar perturbations. Then we discuss these equations in subhorizon and
superhorizon regimes and compare them with those in the $\Lambda$CDM model. Finally, we solve the perturbed equations
by numerical methods for all wave numbers in the range we are concerned with and calculate the integrated Sachs-Wolfe effect
and linear matter power spectrum. We find that the linear matter power spectrum in EiBI gravity deviates from
that in the $\Lambda$CDM model when $k\gtrsim 0.1 h$ Mpc$^{-1}$, which can be further tested in the future by observations
from galaxy surveys.

Arrangement for this paper is as follows. In section \ref{background}, we briefly review the framework of
EiBI gravity and its application to cosmology. In section \ref{perturbation}, we discuss the linear scalar
perturbed equations in EiBI gravity. In section \ref{numerical}, we solve the perturbed equations and
compare the results with those in the $\Lambda$CDM model. In section \ref{obs}, we discuss the integrated
Sachs-Wolfe effect and the linear matter power spectrum in EiBI gravity. Finally, conclusions and
discussions are presented in Section \ref{conclusion}.

\section{Field Equations and Cosmological Background}\label{background}

The action for EiBI gravity is given by \cite{Banados:2010ix}
{
\begin{equation}
S=\frac{2}{\kappa}\int{d^4 x\left[\sqrt{-|g_{\mu\nu}+\kappa R_{\mu\nu}(\Gamma)|}-\lambda\sqrt{-|g_{\mu\nu}|} \right]}+S_{M},
\label{EiBI_Action}
\end{equation}}
where $R_{\mu\nu}(\Gamma)$ represents the symmetric part of the Ricci tensor built with the connection $\Gamma$ and $\lambda$
is a dimensionless constant which is different from $0$ ({here we work in Planck units $c=8 \pi G=1$}). In the following
part we will take $\lambda=1+\kappa \Lambda$, as it is well known that $\Lambda$ here acts as an effective cosmological
constant when $|\kappa R|$ is small \cite{Banados:2010ix}. Varying the action (\ref{EiBI_Action}) independently with respect
to the metric and the connection, respectively, yields
\begin{eqnarray}
\sqrt{q}q^{\mu\nu}&=&\lambda \sqrt{g}g^{\mu\nu}-\kappa\sqrt{g}T^{\mu\nu},\label{Field_Eq_1}\\
q_{\mu\nu}&=&g_{\mu\nu}+\kappa R_{\mu\nu},\label{Field_Eq_2}
\end{eqnarray}
where $q_{\mu\nu}$ is the auxiliary metric compatible with the connection.

Now we consider the case of a homogeneous and isotropic Universe which can be described by the
Friedmann-Robertson-Walker (FRW) metric
\begin{equation}
ds^2=-dt^2+a^2(t)\delta_{i j}dx^i dx^j.
\label{FRW}
\end{equation}
The corresponding auxiliary metric is taken to be
\begin{equation}
q_{\mu\nu}dx^{\mu}dx^{\nu}=-X(t)^2dt^2+a^2(t)Y(t)^2\delta_{ij}dx^i dx^j.
\label{aux_metric}
\end{equation}
For simplicity, we have assumed the spacetime to be spatial flat. Furthermore, we assume that the matter field
is dominated by pressureless cold dark matter and the effect of radiation can be neglected. So the energy-momentum
tensor of matter field can be written as $T_{\mu\nu}=\rho u_{\mu} u_{\nu}$. Then solving
Eqs. (\ref{Field_Eq_1}) and (\ref{Field_Eq_2}) yields
\cite{Banados:2010ix,EscamillaRivera:2012vz,Scargill:2012kg}
\begin{equation}
H^2=\frac{G}{6 F^2},
\label{Friedmann_Eq}
\end{equation}
with
\begin{eqnarray}
G&=&\frac{1}{\kappa}\left(1+2 X^2-3\frac{X^2}{Y^2}\right),\label{G}\\
F&=&1-\frac{3\kappa(1+\kappa \Lambda)\rho}{4[1+\kappa (\rho+\Lambda)](1+\kappa \Lambda)},\label{F}\\
X^2&=&\sqrt{\frac{(1+\kappa \Lambda)^3}{1+\kappa(\rho+\Lambda)}},\label{X2}\\
Y^2&=&\sqrt{[1+\kappa (\rho+\Lambda)](1+\kappa \Lambda)}\label{Y2}.
\end{eqnarray}
Here $\rho$ is the energy density of dark matter.

If $|\kappa|$ is sufficiently small so that $\{|\kappa \rho|,|\kappa\Lambda|\}\ll1$,
Eq. (\ref{Friedmann_Eq}) can be expended in terms of $\kappa \rho$ and $\kappa \Lambda$:
\begin{equation}
H^2=\frac{1}{3}(\rho+\Lambda)+\frac{1}{8}\kappa\rho^2+\mathcal{O}((\kappa\rho)^2).
\label{Friedmann_Eq_O1}
\end{equation}
Note that, since $\rho$ is larger than $\Lambda$ or at least has the same order as $\Lambda$ up to now, we have
used $\mathcal{O}((\kappa\rho)^2)$ to stand for the higher-order terms such as $(\kappa\rho)^2$,
$\kappa^2\rho\Lambda$, $(\kappa\Lambda)^2$, etc. When $\kappa\rightarrow 0$, Eq. (\ref{Friedmann_Eq_O1}) reduces
to the standard Friedmann equation with a cosmological constant. Taking the derivative with respect to $t$ of
Eq. (\ref{Friedmann_Eq_O1}) and considering the continuity equation for cold dark matter $\dot{\rho}+3 H \rho=0$,
we can obtain a useful equation
\begin{equation}
\dot{H}=-\frac{1}{2}\rho-\frac{3}{8}\kappa\rho^2+\mathcal{O}((\kappa\rho)^2).
\label{H_dot}
\end{equation}
It will be used to simplify the perturbed equations in the following sections.

To parametrize Eqs. (\ref{Friedmann_Eq_O1}) and (\ref{H_dot}) for later calculations, we take
\begin{eqnarray}
\rho&=&3 H_0^2 \Omega_m a^{-3},\\
\Lambda&=&3 H_0^2 \Omega_{\Lambda},\\
\gamma&=&\kappa H_0^2.
\label{bg_par}
\end{eqnarray}
Here $\Omega_m$, $\Omega_{\Lambda}$, and $H_0$ are the matter density parameter, the vacuum energy
density parameter, and the Hubble constant at present, respectively ($a$ is normalized so that $a=1$
at present), and $\gamma$ characterizes the deviation from general relativity. These parameters
satisfy
\begin{equation}
\Omega_m+\Omega_{\Lambda}+\frac{9}{8}\gamma \Omega_m^2 = 1.
\label{OmOLG}
\end{equation}
So only two of them are independent. Furthermore, it is also useful to define a dimensionless Hubble parameter
$h$, so that $H_0=100h$ km s$^{-1}$ Mpc$^{-1}$.

\section{Linear Scalar Perturbations}\label{perturbation}

Now we consider a perturbed FRW metric in the Newtonian gauge (here we are only concerned with the
scalar perturbations):
\begin{equation}
ds^2=-(1+2\Phi)dt^2+a^2(t)(1-2\Psi)\delta_{i j}dx^i dx^j.
\label{FRW_pert}
\end{equation}
Noting that the auxiliary metric is related to the physical metric and matter fields according
to Eq. (\ref{Field_Eq_1}), the corresponding perturbed auxiliary metric is taken to be
\begin{equation}
q_{\mu\nu}dx^{\mu}dx^{\nu}=-X(t)^2(1+2\alpha)dt^2+a^2(t)Y(t)^2(1-2\beta)\delta_{ij}dx^i dx^j.
\label{aux_metric_pert}
\end{equation}
The relations between the perturbations of the auxiliary metric and the physical
metric are given in Refs. \cite{Scargill:2012kg,Liu:2012rc,Yang:2013hsa} as
\begin{eqnarray}
\alpha&=&\Phi-\frac{1}{4}\,\frac{\kappa\delta\rho}{1+\kappa(\rho+\Lambda)},\label{alpha}\\
\beta&=&\Psi-\frac{1}{4}\,\frac{\kappa\delta\rho}{1+\kappa(\rho+\Lambda)}.\label{beta}
\end{eqnarray}
Then the equations for the growth of perturbations in the linear regime have been obtained in
Ref. \cite{Yang:2013hsa}.

The perturbed conservation equations in Fourier space lead to
\begin{eqnarray}
\dot{\delta}&=&3\dot{\Psi}+\frac{k^2}{a^2}\delta u,
\label{conser_pert0}\\
\dot{\delta u}\!&=&-\Phi.
\label{conser_perti}
\end{eqnarray}
Here $\delta=\frac{\delta \rho}{\rho}$ is the relative energy density perturbation, and $\delta u$ is related
to the longitudinal part of the spatial velocity perturbation: $\delta u_i^L=\partial_i\delta u$.
The perturbed Eq. (\ref{Field_Eq_2}) gives
\begin{eqnarray}
\!&\!\!&\!\ \frac{X^2}{Y^2}a^{-2}\nabla^2\Phi
     +6\Big(\frac{\ddot{a}}{a}+\frac{\ddot{Y}}{Y}-H\frac{\dot{X}}{X}
          +2H\frac{\dot{Y}}{Y}-\frac{\dot{X}}{X}\frac{\dot{Y}}{Y}\Big)\Phi
     +3\Big(H+\frac{\dot{Y}}{Y}\Big)\dot{\Phi}
     +3\ddot{\Psi}
     +6\Big(H-\frac{1}{2}\frac{\dot{X}}{X}+\frac{\dot{Y}}{Y}\Big)\dot{\Psi}\nonumber\\
\!&\!\!&\!\
     -\frac{1}{4}\kappa a^{-2}\frac{X^2}{Y^2}\frac{\nabla^2\delta\rho}{1+\kappa(\rho+\Lambda)}
     -\frac{3}{4}\kappa \partial_{t}^2\Big[\frac{\delta\rho}{1+(\rho+\Lambda)}\Big]
     -\frac{3}{4}\kappa\Big(3H+3\frac{\dot{Y}}{Y}-\frac{\dot{X}}{X}\Big)
                 \partial_t\Big[\frac{\delta\rho}{1+\kappa(\rho+\Lambda)}\Big]\nonumber\\
\!&\!\!&\!\
     -\frac{1}{2}\Big[1+3\kappa\Big(\frac{\ddot{a}}{a}+\frac{\ddot{Y}}{Y}
          -H\frac{\dot{X}}{X}+2H\frac{\dot{Y}}{Y}
          -\frac{\dot{X}}{X}\frac{\dot{Y}}{Y}\Big)\Big]
          \frac{\delta\rho}{1+\kappa(\rho+\Lambda)}
    -\kappa a^{-2}\partial_t \Big[\frac{\rho}{1+\kappa(\rho+\Lambda)}
          \nabla^2\delta u\Big]\nonumber\\
\!&\!\!&\!\
    -\kappa a^{-2}\Big(2\frac{\dot{Y}}{Y}-\frac{\dot{X}}{X}\Big)
    \frac{\rho}{1+\kappa(\rho+\Lambda)}\nabla^2\delta u=0,
\label{Eins_per00}
\end{eqnarray}
\begin{eqnarray}
\!&\!\!&\!\Big(H+\frac{{\dot Y}}{Y}\Big)\Phi
   +\dot{\Psi}-\frac{1}{4}\kappa\partial_t \Big[\frac{\delta\rho}{1+\kappa(\Lambda+\rho)}\Big]
   -\frac{1}{4}\kappa\Big(H+\frac{\dot{Y}}{Y}\Big)
    \frac{\delta\rho}{1+\kappa(\Lambda+\rho)}
   +\frac{1}{2}\frac{\rho}{1+\kappa(\Lambda+\rho)}\delta u = 0,
\label{Eins_per0i}
\end{eqnarray}
\begin{eqnarray}
\!&\!\!&\!\ \Phi-\Psi=\kappa\frac{Y^2}{X^2}
  \Big[\partial_t\Big(\frac{\rho}{1+\kappa(\rho+\Lambda)}\delta u\Big)
      +\Big(H-\frac{\dot{X}}{X}+3\frac{\dot{Y}}{Y}\Big)
        \frac{\rho}{1+\kappa(\rho+\Lambda)}\delta u \Big].
\label{Eins_perij}
\end{eqnarray}
Substituting Eqs. (\ref{X2}), (\ref{Y2}) into Eqs. (\ref{Eins_per00}), (\ref{Eins_per0i}), (\ref{Eins_perij})
and expanding them with respect to $\kappa\rho$, we obtain
\begin{eqnarray}
\!&\!\!&\!\ 3H(1-\frac{3}{4}\kappa\rho)\dot{\Phi}
   +6\Big[\dot{H}(1-\frac{3}{4}\kappa\rho)+H^2\Big]\Phi
   -\frac{k^2}{a^2}\Phi+3\ddot{\Psi}
   +6H\Big(1+\frac{3}{8}\kappa\rho\Big)\dot{\Psi}\nonumber\\
\!&\!\!&\!\
   -\frac{3}{4}\kappa\rho\ddot{\delta}
   -\frac{3}{4}\kappa\rho H\dot{\delta}
   +\frac{3}{4}\kappa\rho(\dot{H}-2H^2)\delta
  -\frac{1}{2}\rho[1-\kappa(\rho+\Lambda)]\delta
  +\frac{1}{4}\kappa\rho\frac{k^2}{a^2}\delta
 =\mathcal{O}((\kappa\rho)^2),
\label{Eins_per00_1}
\end{eqnarray}
\begin{eqnarray}
   H\Big(1-\frac{3}{4}\kappa\rho\Big)\Phi
  +\dot{\Psi}-\frac{1}{4}\kappa\rho\dot{\delta}
  +\frac{1}{2}\kappa\rho H\delta
  +\frac{1}{2}\rho [1-\kappa(\rho+\Lambda)]\delta u=\mathcal{O}((\kappa\rho)^2),
\label{Eins_per0i_1}
\end{eqnarray}
\begin{eqnarray}
\!&\!\!&\!\ \Phi-\Psi=\kappa\rho(\dot{\delta u}-2 H\delta u)+\mathcal{O}((\kappa\rho)^2).
\label{Eins_perij_1}
\end{eqnarray}
Here we have used the continuity equation $\dot{\rho}+3 H \rho=0$ to eliminate the term $\dot{\rho}$ and
written the equations in Fourier space. Equations. (\ref{conser_pert0}), (\ref{conser_perti}), (\ref{Eins_per00_1}),
(\ref{Eins_per0i_1}), and (\ref{Eins_perij_1}) govern the cosmological scalar perturbations $\Phi$, $\Psi$, $\delta$
and $\delta u$. But it should be noted that only four of theses equations are actually independent. Given appropriate
initial conditions and the expansion history governed by Eqs. (\ref{Friedmann_Eq_O1}) and (\ref{H_dot}),
we can solve these differential equations and calculate related observational quantities to compare
with observations. However, these equations are too complicated for an analytic treatment, so we first
look at two wavelength regimes: wavelengths much smaller than the Hubble horizon (subhorizon), and wavelengths
much larger than the Hubble horizon (superhorizon). It will give us a first impression how EiBI gravity
deviates from general relativity. Then we will show the numerical results in the next section.

\subsection{Subhorizon regime}\label{deep_h}

First, we look at the perturbations which are deep inside the Hubble horizon, i.e., $\frac{k}{a}\gg H$.
To get the equation governing the perturbation $\delta$ deep inside the Hubble horizon, we also apply
the following approximations \cite{Tsujikawa:2007gd,Tsujikawa:2009ku,Thakur:2013oya}:
\begin{equation}
    \Big\{\frac{k^2}{a^2}|\Phi|,\frac{k^2}{a^2}|\Psi|\Big\}
\gg \{H^2|\Phi|,H^2|\Psi|,H|\dot{A}|,|\ddot{A}|\},
\label{approx}
\end{equation}
where $A=\Phi, \Psi$. Then from Eqs. (\ref{conser_pert0}), (\ref{conser_perti}), (\ref{Eins_per00_1}),
and (\ref{Eins_perij_1}), we finally arrive at
\begin{equation}
  \Big(1-\frac{3}{4}\kappa\rho\Big)\ddot{\delta}
 +2H\Big(1-\frac{3}{8}\kappa\rho\Big)\dot{\delta}
 -\frac{1}{2}\Big(\rho-\frac{1}{2}\kappa\rho\frac{k^2}{a^2}\Big)\delta=0.
\label{over_den}
\end{equation}
Note that if $|\kappa\rho|$ is sufficiently small so that the terms associating with it can be neglected except
the term $-\frac{1}{2}\kappa\rho\frac{k^2}{a^2}$ due to large value of $k$, Eq. (\ref{over_den})
is exactly the same as that derived by a different method in the nonrelativistic regime in Ref. \cite{Avelino:2012ge}.

Unlike in general relativity, the pressureless cold dark matter has a nonzero effective sound
speed $c_{seff}=\frac{1}{2}\sqrt{\kappa\rho}$ in EiBI gravity. For positive $\kappa$,
the perturbation of dark matter density exhibits an oscillating behavior when
$\frac{1}{2}\kappa\frac{k^2}{a^2}>1$, which was first pointed out by
Avelino \cite{Avelino:2012ge}. However, in the case we consider in this paper, we will choose
$|\kappa|$ sufficiently small so that $|\frac{1}{2}\kappa\frac{k^2}{a^2}|<1$ for all wave numbers in the
range we are concerned with.

An important observational quantity is the growth rate of clustering defined as
\begin{equation}
f\equiv\frac{d\ln{\Delta}}{d\ln{a}},
\label{growth_f}
\end{equation}
where $\Delta$ is the the relative density perturbation in the comoving gauge,
\begin{equation}
\Delta=\delta-3H\delta u.
\label{D}
\end{equation}
Here we do not use the relative density perturbation in the Newtonian gauge because it depends on the
specific gauge we choose, while the observational quantities should be gauge invariant. It is
easy to check that the combination $\delta-3H\delta u$ is invariant under gauge transformations.

\subsection{Superhorizon Regime}\label{super_h}

Now we go on with the superhorizon regime, i.e., $\frac{k}{a} \ll H$.
It is well known that the quantity $\mathcal{R}\equiv-\Psi+H\delta u$ defined in the Newtonian gauge is
conserved outside the Hubble horizon in general relativity \cite{Bardeen:1980,Lyth:1985}. In Ref. \cite{Bertschinger:2006aw},
Bertschinger had proven that the constancy of $\mathcal{R}$ also holds for modified
gravity theories that obey the energy-momentum conservation $\nabla_\mu T^{\mu\nu}=0$ (see also in Ref. \cite{Hu:2007pj}).
Thus we have
\begin{equation}
-\dot{\Psi}+H\dot{\delta u}+\dot{H}\delta u=0.
\label{conser_out}
\end{equation}
Along with Eq. (\ref{conser_perti}), we can get
\begin{equation}
  \ddot{\Psi}+H\dot{\Phi}-\frac{\ddot{H}}{\dot{H}}\dot{\Psi}
 +H\Big(2\frac{\dot{H}}{H}-\frac{\ddot{H}}{\dot{H}}\Big)\Psi=0.
\label{conser_out1}
\end{equation}
It is the same as the case in general relativity. However, from Eqs. (\ref{conser_perti}), (\ref{Eins_perij_1}),
and (\ref{conser_out}), we can obtain
\begin{equation}
\Psi-\Phi=
\kappa\rho\,\Big[2\frac{H}{\dot{H}}\dot{\Psi}+\Big(1+2\frac{\,\,H^2}{\dot{H}}\Big)\Phi\Big],
\label{conser_out2}
\end{equation}
which implies that $\Phi$ does not equal to $\Psi$ as in general relativity when the matter
fields carry no anisotropic stress. But the deviation will be suppressed by the expansion
of the Universe because $\rho\propto a^{-3}$.

An important quantity is the metric combination
\begin{equation}
\Psi_{+}\equiv\frac{\Psi+\Phi}{2},
\label{psi_plus}
\end{equation}
which affects the CMB power spectrum through the integrated Sachs-Wolfe effect
and weak gravitational lensing. We will discuss the integrated Sachs-Wolfe effect in detail later.

\section{Numerical Evolution}\label{numerical}

In this section, we present the numerical solutions for the scalar perturbations.
We choose Eqs. (\ref{conser_pert0}), (\ref{conser_perti}),
(\ref{Eins_per0i_1}), and (\ref{Eins_perij_1}) to be a complete set of differential equations,
because they are only first order and easier to solve. As mentioned before, we need to give
appropriate initial conditions. It requires a knowledge of the evolution of the perturbations
at early time before the photon decoupled with matter (last scattering). However, to give an accurate prescription
of this era, we need to solve the multispecies Boltzmann equations \cite{Seljak:1996is,Zaldarriaga:1997va,Zaldarriaga:1999ep},
which is beyond the scope of this paper. We will focus on the evolution of linear perturbations from
the time when photon decoupled with matter to the present. At the time of decoupling,
the Universe is already dominated by matter fields. Neglecting the effects of radiation, baryonic matter,
and cosmological constant $\Lambda$, we can solve the perturbed equations analytically. From the
analytic solutions, we assume a set of initial conditions. Using these initial conditions, we
solve Eqs. (\ref{conser_pert0}), (\ref{conser_perti}),(\ref{Eins_per0i_1}), and (\ref{Eins_perij_1})
numerically.

\subsection{Initial Conditions}\label{ini_con}

In section \ref{perturbation}, we have expanded the perturbed equations with respect to $\kappa\rho$.
The zeroth-order equations are just the same as those in general relativity. So we assume the
solutions can be expanded as
\begin{eqnarray}
\delta&=&\delta^{(0)}+\gamma\,\delta^{(1)}+\mathcal{O}(\gamma^2),\label{1order_delta}\\
\Phi&=&\Phi^{(0)}+\gamma\,\Phi^{(1)}+\mathcal{O}(\gamma^2),\label{1order_phi}\\
\Psi&=&\Psi^{(0)}+\gamma\,\Psi^{(1)}+\mathcal{O}(\gamma^2),\label{1order_psi}\\
\delta u\!&=&\delta u^{(0)}+\gamma\,\delta u^{(1)}+\mathcal{O}(\gamma^2),\label{1order_du}
\end{eqnarray}
where $\delta^{(0)}$, $\Phi^{(0)}$, $\Psi^{(0)}$, and $\delta u^{(0)}$ are the same as
the solutions in general relativity ($\kappa\rightarrow0$):
\begin{eqnarray}
\delta^{(0)}&=&2 c_1\Big(1+\frac{k^2}{3\Omega_m H_0^2}a\Big)
   +c_2\Big(3 a^{-\frac{5}{2}}
            -\frac{2 k^2}{3 \Omega_m H_0^2}a^{-\frac{3}{2}}\Big),\label{delta_0}\\
\Phi^{(0)}&=&-c_1+c_2 a^{-\frac{5}{2}}\label{Phi_0},\\
\Psi^{(0)}&=&-c_1+c_2 a^{-\frac{5}{2}}\label{Psi_0},\\
\delta u^{(0)}\!&=&\frac{1}{\Omega_m^{\frac{1}{2}}H_0}
\Big(\frac{2}{3}c_1 a^{\frac{3}{2}}+c_2 a^{-1}\Big).\label{deltau_0}
\end{eqnarray}
We are only interested in the growth modes of
$\delta$ which are important for structure formation, so we take $c_2=0$. Then substituting
Eqs. (\ref{1order_delta}), (\ref{1order_phi}), (\ref{1order_psi}), and (\ref{1order_du}) into
Eqs. (\ref{conser_pert0}), (\ref{conser_perti}), (\ref{Eins_per0i_1}), and (\ref{Eins_perij_1}),
we can obtain the first-order solutions
\begin{eqnarray}
 \delta^{(1)}&=&
  c_1\Big( \frac{3}{4}\Omega_m a^{-3}-\frac{k^2}{2 H_0^2}a^{-2}
           +\frac{k^4}{2\Omega_m H_0^4}a^{-1}\Big)
  +2 c_3\Big(1+\frac{k^2}{3\Omega_m H_0^2}a\Big)
  +c_4\Big(3 a^{-\frac{5}{2}}
           -\frac{2 k^2}{3\Omega_m H_0^2}a^{-\frac{3}{2}}\Big),   \label{delta_1}\\
 \Phi^{(1)}&=&
   -c_1\Big(\frac{3\Omega_m}{4}a^{-3}+\frac{k^2}{4 H_0^2}a^{-2}\Big)
   -c_3+c_4 a^{-\frac{5}{2}},\label{Phi_1}\\
\Psi^{(1)}&=&
  -c_1 \Big(-\frac{\Omega_m}{4}a^{-3}+\frac{k^2}{4 H_0^2}a^{-2} \Big)
  -c_3+c_4 a^{-\frac{5}{2}},\label{Psi_1}\\
 \delta u^{(1)}\!&=&
   \frac{1}{\Omega_m^{\frac{1}{2}}H_0}
   \Big[-c_1\Big(\frac{1}{8}\Omega_m a^{-\frac{3}{2}}
                 +\frac{k^2}{2 H_0^2}a^{-\frac{1}{2}}\Big)
        +\frac{2}{3} c_3 a^{\frac{3}{2}}
        +c_4 a^{-1}
   \Big].\label{deltau_1}
\end{eqnarray}
Is it easy to find that the contributions of the terms associating with $c_3$ and $c_4$ in the above
equations are just modifications to the integral constants $c_1$ and $c_2$ in the
zeroth-order solutions, while the terms associating with $c_1$ present much more
interesting properties. So we assume $c_3=c_4=0$. Actually, if $c_1$, $c_3$, and
$c_4$ all have the same order, the terms associating with $c_1$ will be the dominated
parts of the first-order solutions at the initial time.

Under the above assumptions, we use the analytic forms of the approximate solutions to determine
the initial values of $\delta$, $\Phi$, and $\delta u$. Note that $\Psi$ can be expressed
by $\Phi$ and $\delta u$ algebraically using Eqs. (\ref{conser_perti}) and (\ref{Eins_perij_1}).

\subsection{Numerical Solutions}\label{results}

From Eqs. (\ref{delta_1}), (\ref{Phi_1}), (\ref{Psi_1}), and (\ref{deltau_1}), we can
acquire some important information about the modifications of EiBI gravity to general relativity.
It can be seen that the deviations grow with $k$ at the beginning, but then will be
largely suppressed by the expansion of the Universe. It is also confirmed by our numerical
calculations. As is known that when $k$ is larger, the nonlinear effects will become more significant and
finally make the linear analysis unsuitable. So we restrict $k$ in the range $(0,0.5 h)$Mpc$^{-1}$.
As mentioned in Section \ref{deep_h}, we require $|\frac{1}{2}\kappa\frac{k^2}{a_i^2}|<1$,
where $a_i=\frac{1}{1+z^{*}}$ is the scale factor at the time of decoupling. The redshift $z^{*}$ at
that time is $1090.43\pm0.54$ according to the results of Planck $2013$ \cite{Ade:2013zuv}.
So the parameter $|\gamma|=|\kappa H_0^2|$ should be the order of $10^{-13}$ or smaller.

\begin{figure}[htbp]
\begin{center}
\includegraphics[width=3.3in]{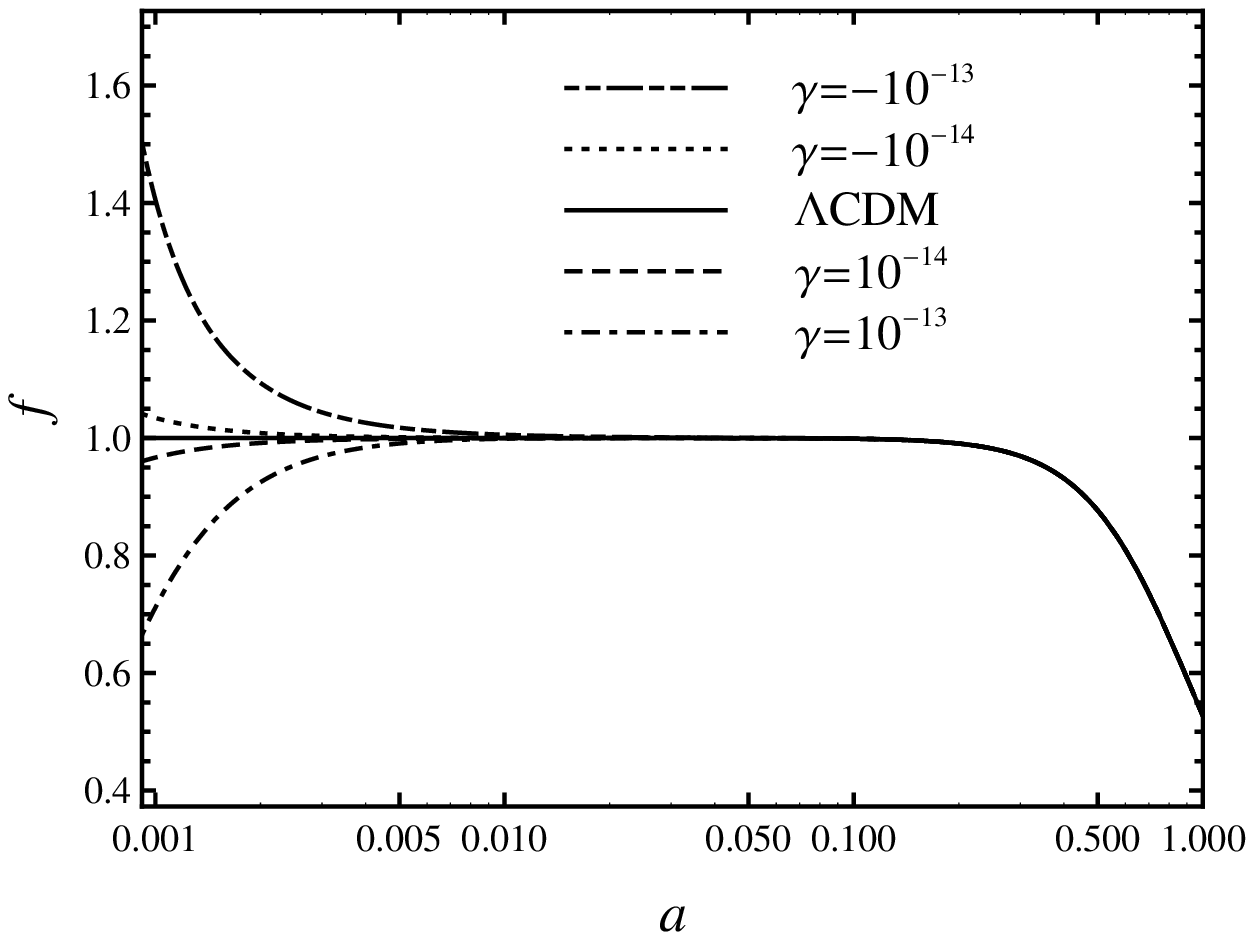}
\includegraphics[width=3.3in]{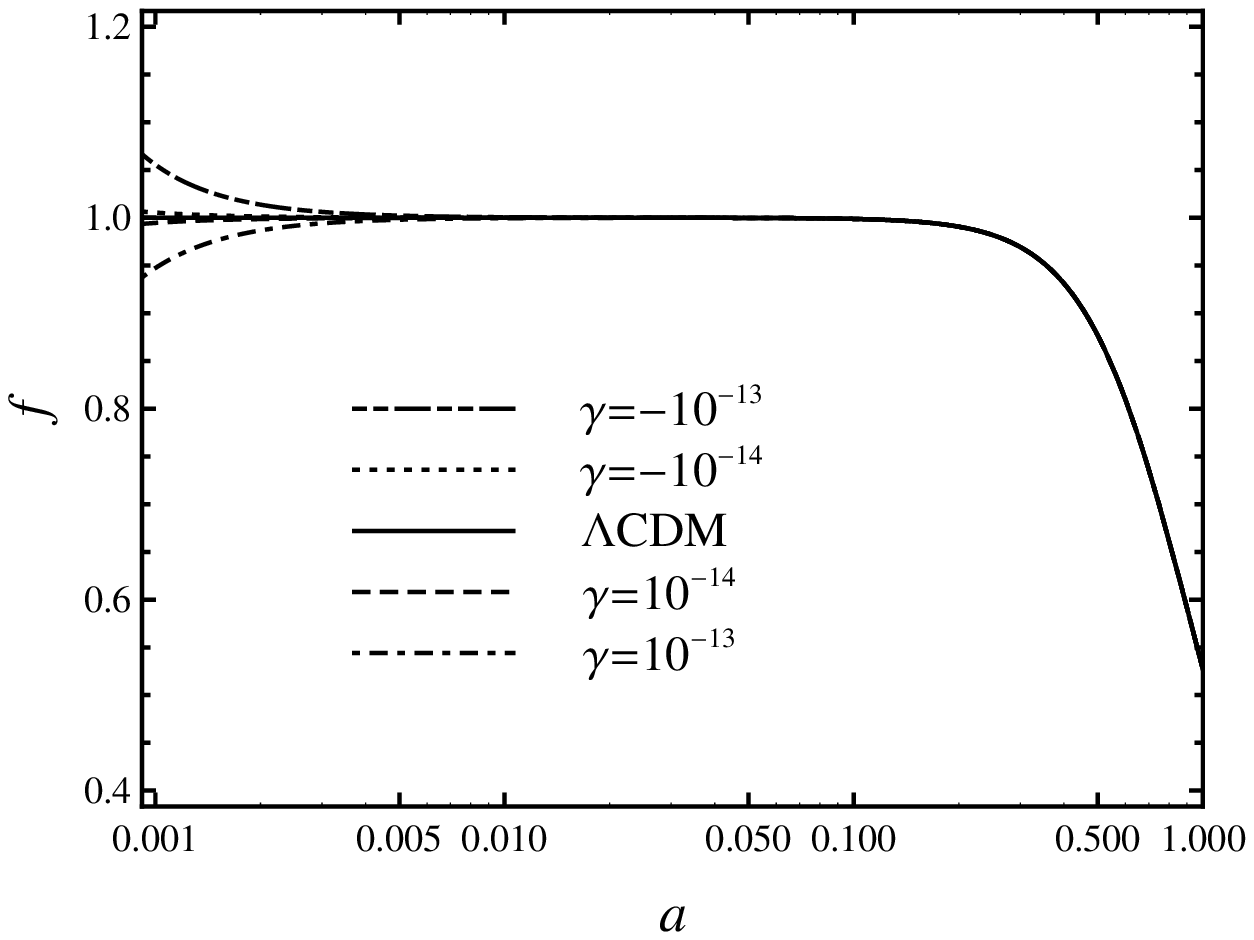}
\end{center}
\caption{Evolution of the growth rate $f$ for $k=0.5 h$Mpc$^{-1}$ (left) and $k=0.2 h$Mpc$^{-1}$ (right)
in EiBI gravity with different values of $\gamma$.}
\label{fig_f}
\end{figure}

Figure \ref{fig_f} shows the growth rate $f$ with respect to the scale factor $a$ for
$k=0.5 h$Mpc$^{-1}$ and $k=0.2 h$Mpc$^{-1}$. The parameter $\gamma$ is taken to be $\pm 10^{-13}$ and
$\pm10^{-14}$. We have also taken $\Omega_m=0.315$ suggested by Ref. \cite{Ade:2013zuv}.
It can be seen that as $|\gamma|$ becomes smaller, the growth rate in EiBI gravity approaches to that in
$\Lambda$CDM model. The deviation is larger for bigger $k$, but due to the expansion of the
Universe, this deviation is largely suppressed at present.
For positive $\gamma$, the growth rate is smaller than that in the $\Lambda$CDM model, so
the growth of structure is suppressed by the effect of the modification to $\Lambda$CDM model at an early time,
which will affect the later process of the formation of the large scale structure. The case with negative $\gamma$
is just the opposite. With the observations from galaxy surveys, we may find some
constraints on EiBI gravity.

\begin{figure}[htbp]
\begin{center}
\includegraphics[width=3.3in]{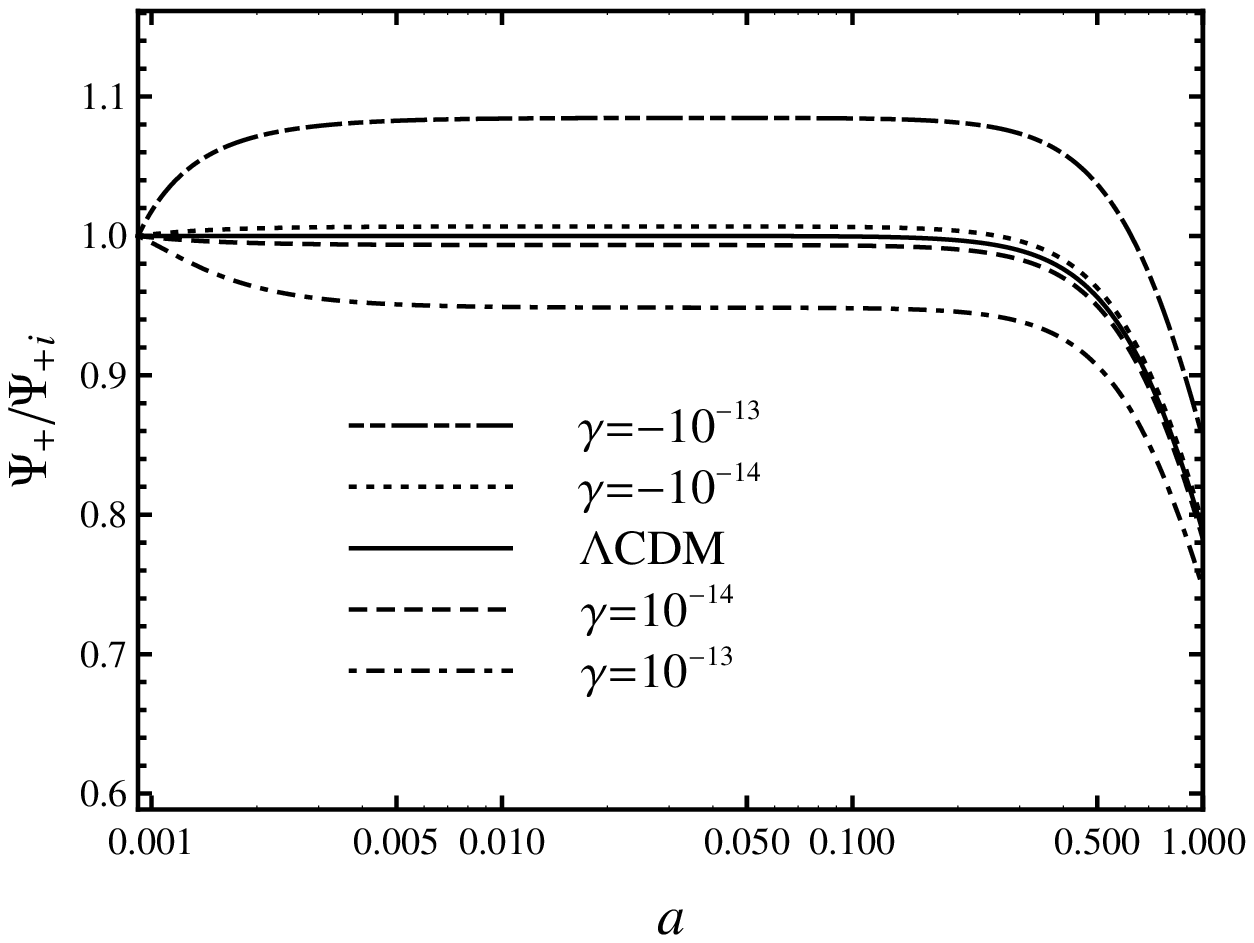}
\includegraphics[width=3.3in]{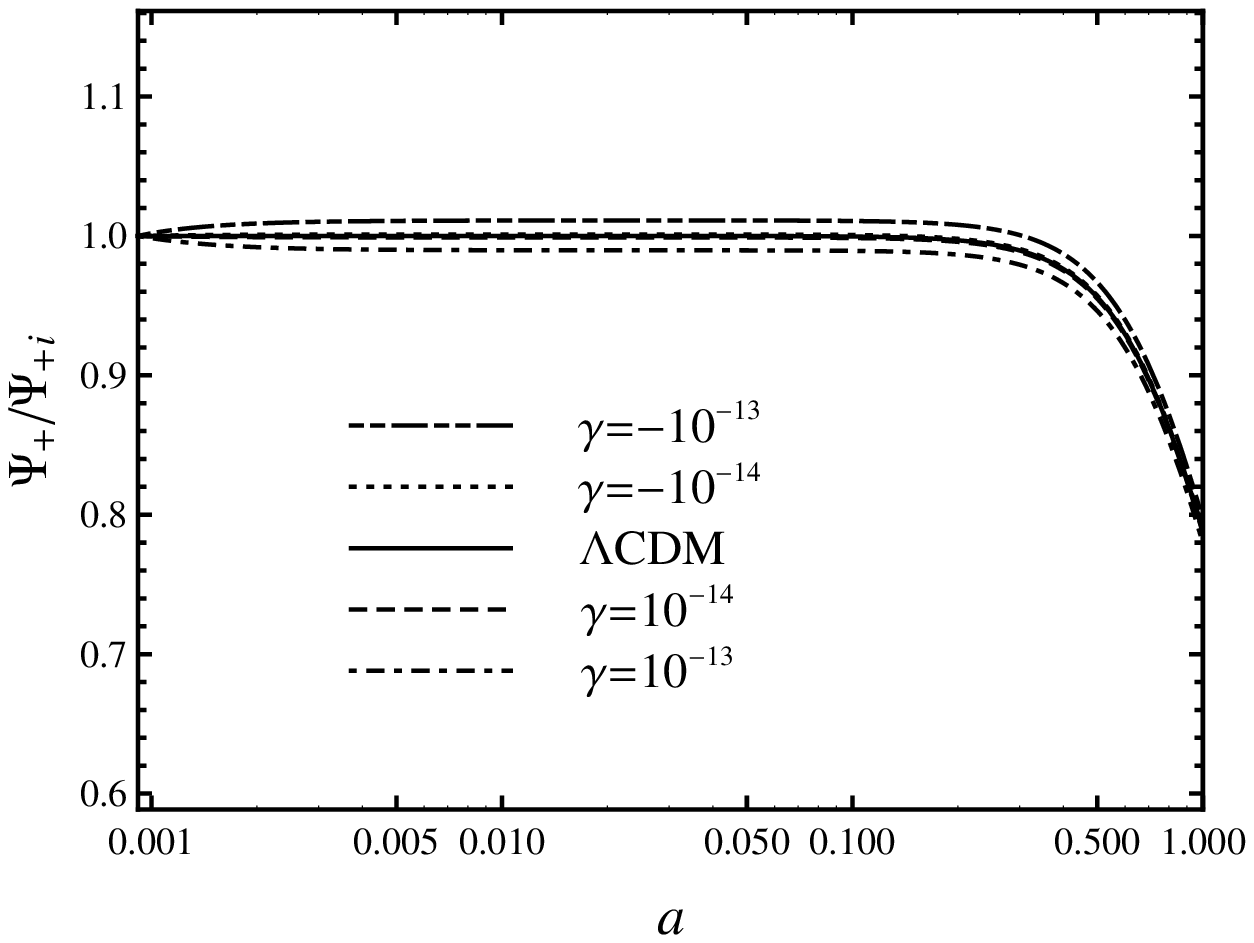}
\end{center}
\caption{Evolution of the metric combination $\Psi_{+}$ for $k=0.5 h$Mpc$^{-1}$ (left) and $k=0.2 h$Mpc$^{-1}$ (right)
in EiBI gravity with different values of $\gamma$. Here $\Psi_{+i}$ is the initial value of $\Psi_{+}$.}
\label{fig_psi_plus}
\end{figure}

Figure \ref{fig_psi_plus} shows the evolution of $\Psi_{+}$ for $k=0.5 h$Mpc$^{-1}$ and $k=0.2 h$Mpc$^{-1}$.
Similarly, as $|\gamma|$ becomes smaller, the evolution of $\Psi_{+}$ in EiBI gravity approaches to that in
$\Lambda$CDM model. The deviation is larger for bigger $k$. The overall change of $\Psi_{+}$ from the
time of decoupling to the present is larger than that in the $\Lambda$CDM model for the case with positive $\gamma$,
while it is the opposite for negative $\gamma$. This deviation from $\Lambda$CDM model will affect the
angular power spectrum of CMB at low multipoles through the integrated Sachs-Wolfe effect.

\section{Observations}\label{obs}

Galaxy surveys, such as the PSCz, 2dF, VVDS, SDSS, 6dF, 2MASS, BOSS, and WiggleZ,
provide us plenty of information about the large scale structure formation. They
report data of the growth rate $f$ at low redshift or its combination with the rms
matter fluctuations at $8h^{-1}$Mpc ($\sigma_8$) and the matter power spectrum
for a large range of $k$. We can use them to test the concordance cosmology model
$\Lambda$CDM and modified gravity theories such as $f(R)$
\cite{Song:2006ej,Lombriser:2010mp,Fu:2010zza,Abebe:2013zua,Hu:2013aqa,Dossett:2014oia}.
Furthermore, WMAP and Planck spacecraft provide us accurate data of the anisotropy of
CMB, which can also be used to constrain different gravity models.

But in our case, according to the results presented in the last section, the growth rate
in EiBI gravity is nearly indistinguishable from that in the $\Lambda$CDM at late time
(low redshift), so we will focus on the effects on CMB and matter power spectrum
below.

\subsection{Integrated Sachs-Wolfe Effect}\label{ISW}

In Ref. \cite{Lagos:2013aua}, the authors showed that we can obtain a nearly scale-invariant
power spectrum for scalar perturbations in EiBI gravity. So we assume the curvature
power spectrum in EiBI gravity can be written as
\begin{equation}
\frac{k^3 P_{\mathcal{R}}}{2\pi^2}=A_s\left(\frac{k}{k_0}\right)^{n_s-1}T(k)^2,
\label{R_spectrum}
\end{equation}
where $A_s$ is the amplitude of curvature power spectrum on the scale $k_0=0.05$Mpc$^{-1}$,
$n_s$ is the scalar spectrum power-law index, and $T(k)$ is the matter-radiation transfer
function.

The integrated Sachs-Wolfe effect contributes to the angular power spectrum of the temperature
anisotropies as \cite{Song:2006ej}
\begin{equation}
C_l^{II}=4\pi\int\frac{dk}{k}[I_l^I]^2\frac{9}{25}\frac{k^3 P_\mathcal{R}}{2\pi^2},
\label{C_ISW}
\end{equation}
where
\begin{equation}
I_l^I(k)=2\int dz G'(z) j_l(k D).
\label{I_l}
\end{equation}
Here $G(z)=\frac{\Psi_{+}(a,k)}{\Psi_{+}(a_i,k)}$, $j_l$ is the spherical
Bessel function, and $D=\int dz/H(z)$ is the comoving distance.

It can be seen that the integrated Sachs-Wolfe effect depends on the variation of
$\Psi_{+}$  with time. For a matter dominated Universe in general relativity, $\Psi_{+}$
is time independent, so there is no integrated Sachs-Wolfe effect. But in EiBI
gravity, even at the matter-dominated era, $\Psi_{+}$ is not time independent.
So there will be a difference between these two theories. As is shown in the last
section, the overall change of $\Psi_{+}$ in EiBI gravity is larger than that
in the $\Lambda$CDM model for positive $\gamma$, while the case with negative $\gamma$
is just the opposite. These deviations will cause an elevation or reduction
of the angular power spectrum at low multipoles.

However, a further analysis shows that the differences caused by the modifications to
general relativity are extremely small. It is not difficult to understand, considering that
the deviations appear for large $k$, but the transfer function $T(k)$ decreases significantly
for these wave numbers, making the contributions from these modes very small.
In fact, if we use the fitting formulas for $T(k)$ proposed in Ref. \cite{Eisenstein:1997ik},
we can calculate the contributions of the integrated Sachs-Wolfe effect. The
results show that the CMB quadrupole power $6 C_2^{II}/2\pi$ contributed by the
integrated Sachs-Wolfe effect is nearly indistinguishable between EiBI gravity
and $\Lambda$CDM model, they all give a value about $362.3$. Here we have also
taken $A_s=2.196\times10^{-9}$, $n_s=0.9603$, and $\Omega_b h^2=0.02205$ \cite{Ade:2013zuv}.
Note that the transfer function is actually different in EiBI gravity. So for a more specific
analysis, we need to calculate it numerically by modifying the numerical codes, such as
CMBFAST \cite{Seljak:1996is,Zaldarriaga:1997va,Zaldarriaga:1999ep} and CAMB \cite{Lewis:1999bs,Howlett:2012mh}.
And a slower but easier to modify program called CMBquick is also available \cite{Creminelli:2011sq}.
But it will be a complicated work and needs a knowledge of the early evolution
of the perturbations at radiation-dominated era and matter-radiation transition in EiBI gravity.

\subsection{Linear Matter Power Spectrum}\label{spectrum}

As in Ref. \cite{Song:2006ej}, we can define the density growth:
\begin{equation}
D_G(a,k)=\frac{\Delta(a,k)}{\Delta(a_i,k)} a_i.
\label{D_G}
\end{equation}
Then the linear matter power spectrum takes the form of
\begin{equation}
\frac{k^3 P_L}{2\pi^2}=\frac{4}{25}D_G^2(a,k)
\frac{k^4}{\Omega_m^2 H_0^4}\frac{k^3 P_{\mathcal{R}}}{2\pi^2}.
\end{equation}
As mentioned in the last subsection, $T(k)$ is different in EiBI gravity. But in this paper,
we focus on the growth of structure after the time of decoupling, thus to compare
with $\Lambda$CDM model we use the same fitting formulate for transfer function as before.

\begin{figure}[htbp]
\begin{center}
\includegraphics[width=3.3in]{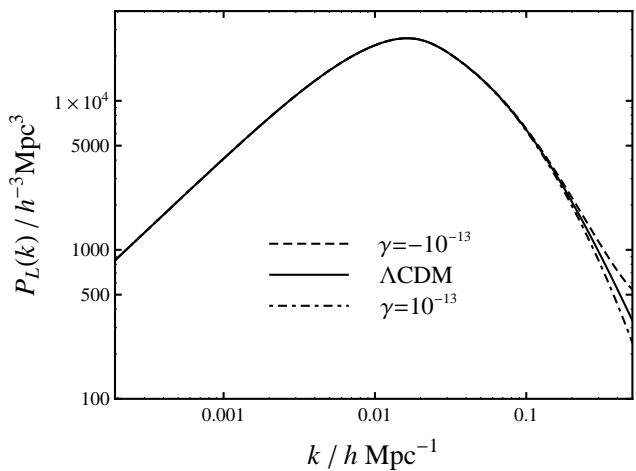}
\end{center}
\caption{Linear matter power spectrum at present in EiBI gravity ($\gamma=\pm 10^{-13}$)
comparing with $\Lambda$CDM model.}
\label{fig_spectrum}
\end{figure}

Figure \ref{fig_spectrum} shows the linear matter power spectrum at present in EiBI
gravity and $\Lambda$CDM model. The parameter $\gamma$ is taken to be $\pm 10^{-13}$.
It can be found that when $k\gtrsim 0.1 h$Mpc$^{-1}$, the power spectrum is lower in
EiBI gravity with positive $\gamma$, while the case with negative $\gamma$ is just the
opposite. It is not a surprising result, if we consider the discussions in section \ref{results}.
For positive $\gamma$, at early time the growth of density perturbation is suppressed by
the modifications of EiBI gravity to $\Lambda$CDM model. This effect is especially
significant for large $k$. On the contrary, for negative $\gamma$, the growth of
density perturbation is strengthened, so the power spectrum is higher
than $\Lambda$CDM at high $k$. Deviations will be much more significant for even larger $k$.

However, with the analysis in this paper, we cannot yet compare the above results
with observations directly. One reason is that we don't know the accurate form of the
transfer function in EiBI gravity. The other is that for large $k$, the nonlinear
effects will become important. So nonlinear analysis is needed in future works.
But from the results of the present work, we can see a trend of increasing departure from
$\Lambda$CDM model with increasing wave number $k$. So there should be a more significant
departure for even larger $k$, when nonlinear regime become important.
Nonlinear measurements of the mass power spectrum through the cluster abundance, Lyman-$\alpha$
forest, and cosmic shear will provide us a way to test EiBI gravity at high $k$ and
give constraint on the only extra parameter $\kappa$.

\section{Conclusions and Discussions}\label{conclusion}

The EiBI gravity has been one of the prospective candidates for modified gravity
theories, in which the singularity at the beginning of the Universe can be avoided. A lot of works have
been done to analyze the stability of cosmological perturbations in this theory and constrain it
using different astrophysical and cosmological observations.

In this paper, we have discussed the evolution of linear scalar perturbations since the matter-dominated era
and the large scale structure formation in EiBI gravity. The growth rate of clustering in EiBI gravity
is found to deviate from that in the $\Lambda$CDM model at an early time of the Universe. The departure increases with wave number $k$.
But at relative low redshift, the growth rate in EiBI gravity approaches to that in the $\Lambda$CDM model.
The suppression (for positive $\gamma$) or enhancement (for negative $\gamma$) on the growth of
density perturbation at early time for large $k$ affects the linear matter power spectrum on small scales (large
$k$). For $k\gtrsim0.1h$Mpc$^{-1}$, the matter power spectrum in EiBI gravity with positive $\gamma$ is lower
than that in the $\Lambda$CDM. And for negative $\gamma$, it is just the opposite. So it is prospective to
use the observational matter power spectrum to test EiBI gravity and constrain the parameter
$\kappa=\gamma/H_0^2$. { If we require that the linear matter power spectrum in EiBI gravity
does not deviate significantly from that in the $\Lambda$CDM model, the parameter $|\gamma|$ should be the
order of $10^{-14}$ or smaller. At present it is still not a very strong constraint.
In Ref. \cite{Avelino:2012qe}, Avelino obtained $|\kappa|<10^{-3}$kg$^{-1}$m$^5$s$^{-2}$ by requiring that
gravity plays a subdominant role inside atomic nuclei, which leads to $|\gamma|<10^{-47}$. If we consider
this constraint, there will be no distinguishable deviation.}
Besides, we also calculate the integrated Sachs-Wolfe effect in EiBI gravity, and find
that its effect on the angular power spectrum of CMB is almost the same as that in the $\Lambda$CDM
model.

However, some work still needs to be done before we can compare the predictions of EiBI gravity directly with
observations. First, we must also analyze in detail the evolution of scalar perturbations at a much
earlier time when the Universe is dominated by radiation, and the transition from a radiation-dominated
Universe to a matter-dominated one. Along with these analyses, we can obtain more accurate transfer function
by modifying corresponding numerical codes such as CAMB, CMBFAST and CMBquick. Secondly, to distinguish
EiBI gravity with $\Lambda$CDM model, we need to compare the evolution of perturbations with large $k$,
at which wave number nonlinear effect cannot be neglected. This can be solved by following the halo-based
description of nonlinear gravitational clustering \cite{Cooray:2002dia} or using numerical simulations.

Although within this paper, we cannot yet give a strong constraint on the parameter $\kappa$ comparable
with that derived from other methods such as considering the compact objects or structure of nucleon,
we have shown some interesting behaviors of EiBI gravity on large scale structure formation, especially
the linear matter power spectrum. An analysis of nonlinear matter power spectrum in future papers
will give a stronger constraint on the deviations from general gravity. Furthermore, despite the
success of $\Lambda$CDM model in predicting the large scale structure, it may have some problems on
small scales, such as the missing satellites problem or the cuspy halo problem. So it is motivated
to consider some modifications on these scales.

\section*{Acknowledgement}

The authors thank Shruti Thakur for her kind reply about the calculation of
some observational quantities. XLD would like to thank Shao-Wen Wei for his
help on figure processing. This work was supported by the National Natural Science
Foundation of China (NSFC) under Grant No. 11375075, and the Fundamental Research
Funds for the Central Universities under Grant No. lzujbky-2013-18.
K. Yang was supported by the Scholarship Award for Excellent Doctoral Student granted
by Ministry of Education. X.H. Meng was partially supported by the Natural Science
Foundation of China (NSFC) under Grant No. 11075078.

\end{document}